\documentclass[
twocolumn,
]{ceurart}

\usepackage{comment}
\usepackage{hyperref}
\usepackage{xcolor}
\usepackage{graphicx}

\begin{document}

\copyrightyear{2021}
\copyrightclause{Copyright for this paper by its authors.
  Use permitted under Creative Commons License Attribution 4.0
  International (CC BY 4.0).}

\conference{5th International GamiFIN Conference 2021 (GamiFIN 2021), April 7-10, 2021, Finland}

\title{Do people's user types change over time? An exploratory study}

\author[1]{Ana Cláudia Guimarães Santos}[%
orcid=0000-0002-3498-0049,
email=anaclaudiaguimaraes@usp.br,
]
\address[1]{Institute of Mathematics and Computer Science, University of São Paulo, São Carlos, São Paulo, Brazil}

\author[1,2]{Wilk Oliveira}[%
orcid=0000-0003-3928-6520,
email=wilk.oliveira@usp.br,
]

\author[2]{Juho Hamari}[%
orcid=0000-0002-6573-588X,
email=juho.hamari@tuni.fi,
]
\address[2]{Gamification Group, Faculty of Information Technology and Communication Sciences, Tampere University,
Tampere, Finland}

\author[1]{Seiji Isotani}[%
orcid=0000-0003-1574-0784,
email=sisotani@icmc.usp.br,
]

\begin{abstract}
  In recent years, different studies have proposed and validated user models (\textit{e.g.}, Bartle, BrainHex, and Hexad) to represent the different user profiles in games and gamified settings. However, the results of applying these user models in practice (\textit{e.g.}, to personalize gamified systems) are still contradictory. One of the hypotheses for these results is that the user types can change over time (\textit{i.e.}, user types are dynamic). To start to understand whether user types can change over time, we conducted an exploratory study analyzing data from 74 participants to identify if their user type (Achiever, Philanthropist, Socialiser, Free Spirit, Player, and Disruptor) had changed over time (six months). The results indicate that there is a change in the dominant user type of the participants, as well as the average scores in the Hexad sub-scales. These results imply that all the scores should be considered when defining the Hexad's user type and that the user types are dynamic. Our results contribute with practical implications, indicating that the personalization currently made (generally static) may be insufficient to improve the users' experience, requiring user types to be analyzed continuously and personalization to be done dynamically.
\end{abstract}


\maketitle

\section{Introduction}
\label{introduction}

In the last decades, video games have become an important part of people's life and culture \cite{ryan2006motivational,hogberg2019gameful,hassan2020gameful} and the positive experience people have playing games can affect positively their behavior \cite{hogberg2019gameful,hassan2020gameful}. To evoke similar positive game experiences, gamification\footnote{In this study defined as ``the use of game elements in non-context games'' \cite{deterding2011game}.} has been used in different contexts including education \cite{roosta2016personalization,araya2019does}, public administration \cite{hassan2017governments,harviainen2019governmental}, and health \cite{johnson2016gamification, orji2018personalizing}. Thus, one of the main goals of gamification is to change people's behavior towards a positive performance in general activities \cite{deterding2011gamification,bai2020gamification}, making the analysis of the users' experience in gamified settings one of the main concerns of researchers in recent years \cite{johnson2016gamification,koivisto2019rise, bai2020gamification}. Therefore, several studies have been conducted to analyze the individual experiences of the participants in gamified settings \cite{mora2019quest,oliveira2020does,altmeyer2020using}. Studies that emphasize users' experience focus on proposing user models according to the individualities of each person \cite{bartle1996hearts,nacke2011brainhex,tondello2019empirical} or personalizing the systems according to these user models \cite{orji2014modeling,lavoue2018adaptive,oliveira2019tailored}.

For many years, research has indicated that people have different opinions and perceptions about everyday things \cite{lockhart1979behavioral,little2011human,palmer2013visual}. The same occurs concerning games and gamification, with several studies indicating that people have different preferences and perceptions about gamification design \cite{orji2018personalizing,orji2017improving,oliveira2019tailored}. Thus, different user models have been proposed to be validated not only for games \cite{bartle1996hearts,yee2006motivations,nacke2011brainhex} but also for the gamification's field  \cite{tondello2019empirical}. Despite the growing number of researches about the personalizing of gamified environments \cite{klock2020tailored,stuart2020tailor}, the results are still contradictory, and not always, the user models are faithful to each person's preferences, as well as, not always, personalization has positive effects on the users' experience \cite{klock2020tailored}. Some recent studies about personalized gamification also considered personality traits to define the preference for gamification designs \cite{hallifax2019factors}. Studies about the stability of personality traits have been conducted to investigate how stable the personality can be over time \cite{cobb2012stability,rantanen2007long,costa1986personality,hampson2006first}, and some of their results showed that some personality traits can be more stable over the years or according to the gender \cite{rantanen2007long}. The studies about personality traits have an impact on the studies about models proposed to define players and user types since the personality traits are related to the player type itself \cite{tondello2019empirical,akgun2018adaptation}.

Theoretical studies argue that one of the reasons for the contradictory results in the researches about the personalizing of gamified environments is that users' preferences may change over time \cite{busch2016player,klock2020tailored}, and thus, user models would need to be dynamic, as well as the personalization of systems \cite{oliveira2019tailored}. However, in general, studies on this field are theoretical and few empirical studies have been carried out until today \cite{klock2020tailored,busch2016player}, not allowing to know if the user types change over time in the gamification context. To start to face the challenge of understating better the user types in gamified settings, in this study, we aim to \textbf{identify if people's user types change over time}, thus answering the question ``Does people's user types (Achiever, Philanthropist, Socialiser, Free Spirit, Player, and Disruptor) change over time (six months)?''. 

To achieve our aim, we conducted an exploratory study with 74 participants where we compared the user types of participants in the first phase with the user types of the same participants in the second phase (six months after). The main results indicate that most of the participants presented different dominant user types (\textit{i.e.}, the strongest tendency of the participants) and also the scores of both phases presented differences. Thus, the obtained results showed that the user types can not be considered stable after six months.

The results obtained have theoretical and practical implications in the design of gamified environments, indicating that the user types of participants possibly change over time. Therefore it is not enough to analyze the participant's user types once, as well as that the static personalization of the gamification might not be enough to provide a good experience for users, being necessary to invest in approaches to provide dynamic personalization, \textit{i.e.}, that changes according to the changing profile of users.

\section{Background}
\label{background}

This section presents the study background (\textit{i.e.}, player and user types), as well, the related works.

\subsection{Player and user types}

Several studies have proposed how users can be disposed of into different player/user types according to their behavior or motivations \cite{bartle1996hearts,nacke2011brainhex,yee2006motivations,marczewski2015even}. The player typology became an important part of studies about gamification design \cite{klock2020tailored} since it helped researchers to understand more how the player/user type can affect the use of gamified systems \cite{tondello2019empirical,hallifax2019factors,oliveira2020does}. The typology model proposed by Bartle \cite{bartle1996hearts} was one of the first and most used player type models \cite{tondello2016gamification, tondello2019empirical}, which classifies the users into four different players: Achievers, Explorers, Killers, and Socializers. This player type model was based on the characteristics of players from multi-user dungeons (MUDs).

Inspired by Bartle's player type model, Yee \cite{yee2006motivations} created an empirical model for player motivations with data collected from Massive Multiplayer Online Role-Playing Games (MMORPGs). Yee \cite{yee2006motivations} identified ten subcomponents categorized into three main components: \textit{i)} Achievement (advancement; mechanics and competition); \textit{ii)} Social (socializing, relationship, and teamwork), and \textit{iii)} Immersion (discovery, role-playing, customization, and escapism). More recently, based on neurobiological findings and data collected from players of different games, Nacke \textit{et al}. \cite{nacke2011brainhex} created the BrainHex with seven archetypes of players (Achiever, Survivor, Conqueror, Mastermind, Seeker, Daredevil, and Socialiser).

Ultimately, Marczewski \cite{marczewski2015even} proposed the Gamification User Types Hexad \cite{marczewski2015even} that classifies the users of gamified environments into six different user types, according to their motivations. This model is the first created specifically for gamification and it was based on the self-determination theory (SDT), which classifies motivation in intrinsic or extrinsic \cite{deci1985conceptualizations}. The user types of the Hexad classified according to intrinsic motivations are the Achiever (motivated by mastery); Philanthropist (motivated by purpose); Socialiser (motivated by relatedness); and Free Spirit (motivated by autonomy) \cite{marczewski2015even,diamond2015hexad,tondello2016gamification}. Player (motivated by rewards) is the user type classified according to extrinsic motivations and the Disruptor (motivated by change) is a user type derived from user behavior observation in online systems \cite{marczewski2015even,diamond2015hexad,tondello2016gamification,ooge2020tailoring}. People show a stronger tendency (\textit{i.e.}, dominant user type) for one of the Hexad user types, and yet they are also motivated by all the other user types in a certain degree \cite{tondello2019empirical}. Thus, the Hexad model presents the user type as a collection of six scores \cite{tondello2019empirical}.

The Hexad have been chosen for this study, considering that is a validated user typology for gamification and have been used in a considerable number of studies about gamification in the last years \cite{orji2018personalizing,mora2019quest,lopez2019effects,altmeyer2020using,hallifax2019factors}.

\subsection{Related works}
\label{sec:related-works}

Recently several studies have been conducted to understand more how the user types could be used in the gamification domain, especially in the context of personalization based on the user type choices. We defined our related works based on the results of different secondary studies on gamification, user types, and personalization \cite{koivisto2019rise,klock2020tailored, bai2020gamification}, as well as, based on a snowballing review. Hallifax \textit{et al}. \cite{hallifax2019factors} conducted a study with 300 participants seeking to identify which player typology would be the most suitable for tailored gamification and the motivational impact of game elements. Considering only the research about user typology, they sought to answer if the dominant user type is sufficient to discriminate users’ preferences and which typology should be chosen for tailored gamification, considering the BrainHex player typology \cite{nacke2014brainhex}, Gamification User Types Hexad \cite{marczewski2015even} and Big Five Factors \cite{goldberg1992development}. Although they have done important findings about user typology (\textit{e.g.}, Hexad is the most appropriate user typology for tailoring gamification \cite{hallifax2019factors}), they did not conduct any research to identify if there were changes in participants' user type over time.

Lopez \textit{et al}. \cite{lopez2019effects} conducted a study with 30 participants to explore the effects of the user type on the users’ performance in a gamified application. To access the user type of the participants, they used the Gamification User Types Hexad \cite{marczewski2015even} and the environment chosen to conduct the study was an application that promotes physical activity. Their findings (\textit{e.g.}, the indication that user type is related to the users’ performance) showed the importance of accessing the user type to increase the chances of success of an environment, however, they did not conduct another study to know if the users’ type had changed over time, and consequently if their performance were the same.

Tondello \textit{et al}. \cite{tondello2019empirical} conducted three large studies (n1=556, n2=1328, n3=152) to validate the scale in English and Spanish of the Gamification User Types Hexad. Their results made possible the validity of the scale in both languages, they correlated some user types (\textit{e.g.}, there is a strong correlation between the Philanthropist and Socialiser user types) and also correlated user types with gender and age (\textit{e.g.}, women were related with the user types with intrinsic motivations). Although the participants of the third study were the same as those in the second study, they did not analyze if the user types were the same, even though it passed six months between the studies.

Altmeyer \textit{et al}. \cite{altmeyer2020hexarcade} conducted a study with 147 participants to analyze the potential of two gameful applications to predict the Hexad user types. In one application, the respondents were asked to select one statement and then they received gameful feedback. On the other, participants first experienced game elements and then decided with which they wanted to interact. Their findings (\textit{e.g.}, participants mostly interacted with game elements that have been related to their user types in other studies) advance the literature showing that might the access of the user type can occur without the use of questionnaires, but they did not consider if the user types would change over time. 

Busch \textit{et al}. \cite{busch2016player} conducted two online studies, (n1=592 and n2=243), about the BrainHex Model \cite{nacke2011brainhex} in the game context. They tried to confirm the validity of the scale and if the respondents' player type were the same after six months. They did confirm that the player type of the respondents was not stable, but they used a game based player typology (BrainHex Model), which may prevent the results to be the same in the context of gamification, and did not present any demographic information that could influence the changing of the player type. Thus, as far as we know, our study is the first study conducted to identify if the user types can change over time, using a validated user type framework to the gamification context. \autoref{tab:related-works} presents a comparison between the related works.

\begin{table*}[!ht]
\caption{Related works comparison}
\label{tab:related-works}
\begin{tabular}{lccccccc}
\hline
\multicolumn{1}{c}{\textbf{Authors}}     & \multicolumn{1}{c}{\textbf{Year}}    &\multicolumn{1}{c}{\textbf{Player/User Typology}}       & \multicolumn{1}{c}{\textbf{AUP}}               & \multicolumn{1}{c}{\textbf{AUS}}                                \\ \hline
Hallifax \textit{et al}. \cite{hallifax2019factors}         &2019  & BrainHex and Hexad                                 &   $\bullet$                                       &                                                                                                             \\ \hline
Lopez \textit{et al}. \cite{lopez2019effects}     &2019  & Hexad     &      $\bullet$              &                                                                                                                 \\ \hline
Tondello \textit{et al}. \cite{tondello2019empirical}    &2019  & Hexad   & $\bullet$          &                                                                                                                   \\ \hline
Altmeyer \textit{et al}. \cite{altmeyer2020hexarcade}    &2020 &  Hexad                                                    & $\bullet$                                   &                                                                                                                   \\ \hline

Busch \textit{et al}. \cite{busch2016player}       &2016 & BrainHex & $\bullet$        &   $\bullet$                                                  
                                             \\ \hline
Our study                 &2020  & Hexad               & $\bullet$              &   $\bullet$                                 
                                             \\ \hline
\multicolumn{6}{p{9cm}}{\textbf{Key}: AUP: Accessed the user type of the participants; AUS: Analyzed if the player/user type is stable over time.} \\ \hline
\end{tabular}
\end{table*}

\section{Data and methods}
\label{sec:data-methods}

The study was organized in two different phases (two interventions), the first one was conducted in March/April 2020, and the second in October 2020. The first phase was divided into: \textit{i)} survey design; \textit{ii)} pilot study; \textit{iii)} survey application; and \textit{iv)} data analysis. In the second phase, we replicated the survey with the same participants and re-analyzed the data. Figure 1 summarizes the study design.

\begin{figure}[!ht]
\label{fig:diagrama}
  \centering
  \includegraphics[width=\linewidth]{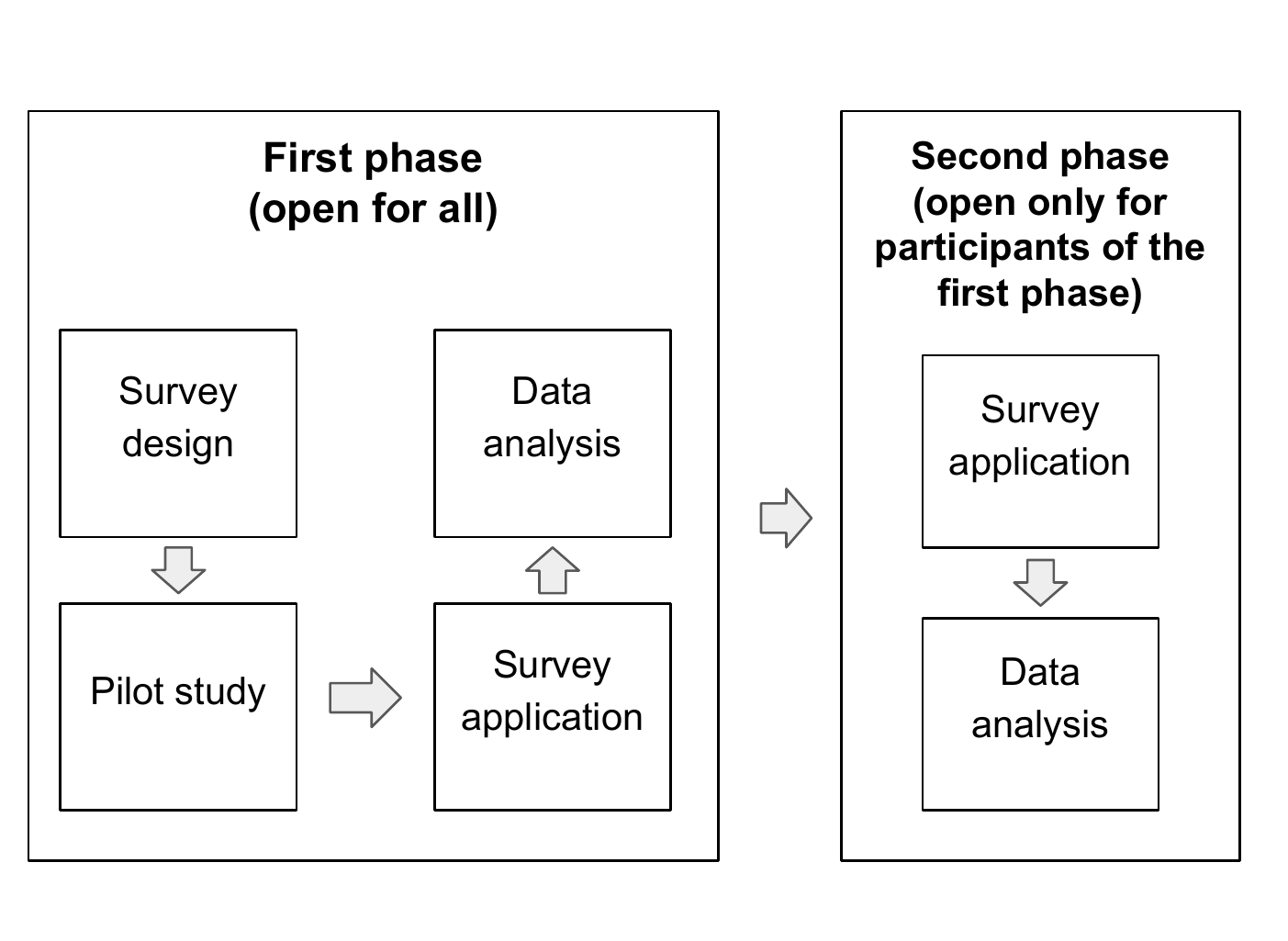}
  \caption{Study design}
\end{figure}

\subsection{Materials}
\label{sec:materials}

In both phases of this study, the questionnaire consisted of 32 questions divided into two different sections. The first one, inspired by previous studies \cite{nacke2014brainhex,tondello2019empirical,oliveira2020does}, was used to collect the participants' demographic information and gaming habits. In this section, the participants could leave their e-mails for future studies, however, this question was not an obligatory question. The second section was used to collect the user type of the participants, using the Hexad scale proposed by Tondello \textit{et al}. \cite{tondello2016gamification}, which consists of 24 questions, evaluated by the respondents in a 7-point Likert Scale \cite{likert1932technique}, and randomly arranged in the questionnaire, preventing the respondents to identify the questions that were similar and compose the user type characteristics, as recommended by Tondello \textit{et al}. \cite{tondello2016gamification}. 

To improve the survey reliability, inspired in other recent studies \cite{orji2018personalizing,hallifax2019factors,oliveira2020does,ooge2020tailoring} we also inserted an ``attention check question''\footnote{``I like to be with my friends, but this question is just to evaluate your attention: please check option 3 to let us know that you are paying attention.''}. The main objective of this question was to guarantee that the respondents were paying attention while answering the questionnaire. All the answers that did not check the right option in the ``attention check question'' were eliminated before the data analysis.

After the construction of the questionnaire in the first study, as recommended by Connelly \cite{connelly2008pilot}, we conducted a pilot study. Our pilot study was conducted with ten respondents, that analyzed if the size of the questionnaire was appropriate. Eight respondents answered that the questionnaire wasn't large, then any question was taken away from it. To perform the statistical analysis we used the software SPSS (Statistical Package for the Social Sciences) 26.

\subsection{Participants}
\label{sec:participants}

For the first phase, the questionnaire was released on March 26th, 2020, was opened for thirty-eight days, and was spread by social networks (Facebook, Twitter, and Instagram) and e-mail (public lists from universities). Were collected 366 answers, of which 331 were valid according to the attention-check question. From these 331 answers, 182 respondents provided a valid e-mail and authorized the contact for other studies. These 182 e-mails were provided by 90 people that self-reported as women (49\%) and 92 people that self-reported as men (51\%). 71\% of the respondents reported that playing games were a habit. 56\% of the respondents presented only one of the six Hexad's user types (Achiever, Philanthropist, Socialiser, Disruptor, Free Spirit, and Player) as dominant user type, while the other 44\% presented twenty-two different combinations of the six Hexad's user types as dominant user type (\textit{e.g.}, Achiever and Philanthropist, Philanthropist and Socialiser). 

The questionnaire of the second phase was released on October 7th, 2020, was opened for twenty-five days, and was sent to all the 182 respondents that left a valid e-mail in the first phase. Were collected 87 answers, of which 74 were valid according to the attention-check question. In this phase, 57\% of the respondents presented only one of the six Hexad's user types (Achiever, Philanthropist, Socialiser, Disruptor, Free Spirit, and Player) as dominant user type, while the other 43\% presented twenty-two different combinations of the six Hexad's user types as dominant user type (\textit{e.g.}, Achiever and Philanthropist, Philanthropist and Socialiser). \autoref{tab:demographic} presents the demographic information and gaming habits of the respondents that participated in both phases of the study.

\begin{table*}[!ht]
  \caption{Demographic information and gaming habits of the participants of both phases}
  \label{tab:demographic}
\begin{tabular}{cccccc}
\hline
\multicolumn{6}{c}{\textbf{Demographic information}}     \\ \hline
\multirow{2}{*}{Gender}          & Female                      & 55\%                  & \multirow{10}{*}{Age}      & 15-19         & 5\%  \\ \cline{2-3} \cline{5-6} 
                                 & Male                        & 45\%                  &                            & 20-24         & 12\% \\ \cline{1-3} \cline{5-6} 
\multirow{8}{*}{Education level} &                             &                       &                            & 25-29         & 16\%  \\ \cline{5-6} 
                                 & Basic Education           & 4\%                   &                            & 30-34         & 14\% \\ \cline{2-3} \cline{5-6} 
                                 & Bachelor                    & 27\%                  &                            & 35-39         & 14\% \\ \cline{2-3} \cline{5-6} 
                                 & Specialized courses         & 27\%                  &                            & 40-44         & 18\% \\ \cline{2-3} \cline{5-6} 
                                 & M.Sc.                       & 26\%                  &                            & 45-49         & 11\% \\ \cline{2-3} \cline{5-6} 
                                 & PhD                         & 9\%                  &                            & 50-54         & 5\%  \\ \cline{2-3} \cline{5-6} 
                                 & PostDoc                     & 7\%                   &                            & 55-59         & 4\%  \\ \cline{5-6} 
                                 &                             &                       &                            & Over 60       & 1\%  \\ \hline
\multicolumn{6}{c}{\textbf{Gaming habits}}                                                                                                                                                  \\ \hline
\multicolumn{2}{c}{Play games}     & \multicolumn{1}{c}{72\%} & \multicolumn{2}{c}{Do not play games}       & \multicolumn{1}{c}{28\%} \\ \hline
\multicolumn{1}{c}{\multirow{4}{*}{Frequency}} & \multicolumn{1}{c}{Everyday}      & \multicolumn{1}{c}{14\%}  & \multicolumn{1}{c}{} & \multicolumn{1}{c}{} & \multicolumn{1}{c}{}   \\ \cline{2-2}
\multicolumn{1}{c}{}                           & \multicolumn{1}{c}{Every week}    & \multicolumn{1}{c}{16\%}  & \multicolumn{1}{c}{} & \multicolumn{1}{c}{} & \multicolumn{1}{c}{}   \\ \cline{2-2}
\multicolumn{1}{c}{}                           & \multicolumn{1}{c}{Rarely}        & \multicolumn{1}{c}{54\%}  & \multicolumn{1}{c}{} & \multicolumn{1}{c}{} & \multicolumn{1}{c}{}   \\ \cline{2-2}
\multicolumn{1}{c}{}                           & \multicolumn{1}{c}{I do not know} & \multicolumn{1}{c}{16\%}  & \multicolumn{1}{c}{} & \multicolumn{1}{c}{} & \multicolumn{1}{c}{}   \\ \hline
\end{tabular}
\end{table*}

To access the dominant user type of the participants in both phases, we measured all the scores that each participant presented in each sub-scale (\textit{i.e.}, the four questions of each Hexad user type) and calculated where they presented the higher average score (\textit{e.g.}, if the respondent presented as score \textbf{28 in the Achiever sub-scale}; 21 in the Philanthropist sub-scale; 20 in the Socialiser sub-scale; 18 in the Player sub-scale; 16 in the Free Spirit sub-scale; and 15 in the Disruptor sub-scale, his/her dominant user type is the Achiever). This way to calculate the dominant user type has been used similarly before \cite{hallifax2019factors,akgun2018adaptation}. Instead of grouping the respondents only in the six Hexad user types, following the original Hexad results \cite{Hexadresults}, the respondents that presented the higher average score repeated in more than one sub-scale were grouped into the combination of all the users they presented as the dominant user type (\textit{e.g.}, if the respondent presented as score \textbf{28 in the Achiever sub-scale; 28 in the Player sub-scale}; 20 in the Philanthropist sub-scale; 20 in the Socialiser sub-scale; 16 in Free Spirit sub-scale; and 15 in Disruptor sub-scale, his/her dominant user type is the Achiever and Player).

The participation in the pilot study and both research phases was voluntary, thus, was not offered to the respondents any remuneration or gifts. Since the size of the questionnaire and the quality of the answers was a concern before the data collection, we understand that, as volunteers, respondents are more willing to pay attention when answering the questionnaire. Tondello and Nacke \cite{tondello2020validation} also presume that volunteers participate in studies without pressure to maximize time usage, which can provide more quality of the answers.

Considering the type of this study, the sample size of both phases is adequate, according to Bentler and Chih-Ping \cite{bentler1987practical} and Hair \textit{et al}. \cite{hair1998multivariate}, which define the necessity of at least five participants for each construct measured. Construct is a latent concept that can be defined in conceptual terms but cannot be directly measured \cite{hair1998multivariate} (\textit{e.g.}, extroversion, autonomy), thus its measurement occurs indirectly through other variables.

\section{Results}
\label{results}

Initially, to ensure the instrument validation for the study, we measured the internal reliability for each Hexad user type in both phases of the research. Overall, the reliability was acceptable ($\alpha$ $\geq$ 0.70, RHO A $\geq$ 0.70, CR $\geq$ 0.70, AVE $\geq$ 0.50) for all user types, except for the user type Disruptor (in both phases) and Free Spirit (first phase), which were slightly below the acceptable. Cronbach's values under 0.7 for the Disruptor and Free Spirit scales were also found in other recent studies \cite{tondello2019empirical,ooge2020tailoring}. The reliability results can be seen in \autoref{tab:reliability}.

\begin{table}[!ht]
\caption{Reliability results}
\label{tab:reliability}
\begin{tabular}{lllll}
\hline
\textbf{Construct}       & \textbf{$\alpha$}      & \textbf{RHO}      & \textbf{CR}      & \textbf{AVE}      \\ \hline 

Achiever1          & \textbf{0.863}                 & \textbf{0.924}                  & \textbf{0.874}        & \textbf{0.642}           \\ \hline
Achiever2         & \textbf{0.845}              &  \textbf{1.148}    &\textbf{0.883}   &  \textbf{0.655}    \\ \hline
Disruptor1       &  \textcolor{gray}{0.648 }         & \textcolor{gray}{0.668}       &    \textbf{0.790}                 &   \textcolor{gray}{0.486}     \\ \hline
Disruptor2     &  \textcolor{gray}{0.660}            &\textcolor{gray}{0.691}       &   \textbf{0.794}                  & \textcolor{gray}{0.494}           \\ \hline
Free Spirit1        & \textcolor{gray}{0.676}      &    \textbf{0.987}    &    \textbf{0.758}      &   \textcolor{gray}{0.451}     \\ \hline
Free Spirit2        & \textbf{0.785}        &  \textbf{0.813}         &    \textbf{0.853}               & \textbf{0.593}                \\ \hline
Philanthropist1         & \textbf{0.873}        &  \textbf{0.987}          &   \textbf{0.909}              &   \textbf{0.716}            \\ \hline
Philanthropist2          &  \textbf{0.890}       &   \textbf{0.914}      &  \textbf{0.924}                 &   \textbf{0.754}                \\ \hline
Player1        & \textbf{0.743}          &   \textbf{0.813}    &     \textbf{0.830}            &  \textbf{0.552}                     \\ \hline
Player2          & \textbf{0.851}         & \textbf{0.855}    &   \textbf{0.900}                   &   \textbf{0.694}                  \\ \hline
Socialiser1             &\textbf{0.862}     &\textbf{0.987}        &   \textbf{0.903}               &  \textbf{0.702}                \\ \hline
Socialiser2             & \textbf{0.883}      &   \textbf{0.893}    &    \textbf{0.919}                &   \textbf{0.740}     \\ \hline
\multicolumn{5}{p{6.5cm}}{\textbf{Key}: $\alpha$: Cronbach's; RHO: Jöreskog’s rho; CR: Composite Reliability; AVE: Average Variance Extracted; 1: results of the first research phase; 2: results of the second research phase; Values in grey are $\alpha$ < 0.70, RHO A < 0.70, CR < 0.70, AVE < 0.50} 
\\ \hline
\end{tabular}
\end{table}

Analyzing the dominant user types it was possible to discover that 57 people (76\% of the participants) presented a change of the dominant user type after six months. Once more, we considered as dominant user type the six Hexad user types and the combination between them that the respondents presented as the higher average score. The 74 participants presented, in the first phase of the study, 19 different user types combinations as dominant user type, with 51\% presenting a combination of more than one user type as the strongest tendency. In the second phase, the 74 participants presented 20 different user types, with 43\% presenting a combination of more than one user type. Females (78\%) changed the strongest tendency more than males (73\%), as well as people who play (77\%) changed more than people who do not play (71\%). People with a master's degree (32\%) were those who more changed the dominant user type in the educational level group. Considering only the age, six of the ten age groups measured in this research, presented a change of more than 75\% (15-19 years - 100\%; 20-24 years - 89\%; 25-29 years - 92\%; 30-34 years - 80\%; 40-44 years - 92\%; 50-54 years - 75\%).

\autoref{tab:hexad1} presents the different combinations that the 74 respondents presented as dominant user types in both phases of the study. To measure if there was a consensus between the dominant user types of both phases, we used the Kappa coefficient that is used to measure the reliability between pairwise agreement \cite{mchugh2012interrater}. The Kappa results vary from -1 to +1 \cite{mchugh2012interrater}, where $\kappa$<0.00 the agreement is considered poor; when $\kappa$ is between 0.00-0.020 the agreement is considered slight; when $\kappa$ is between 0.21-0.040 the agreement is considered fair; when $\kappa$ is between 0.41-0.060 the agreement is considered moderate; when $\kappa$ is between 0.61-0.080 the agreement is considered substantial; and when $\kappa$ is between 0.81-1 the agreement is considered almost perfect \cite{landis1977measurement}. Our result ($\kappa$ = 0,00) indicates that there was a very low agreement between the first and second research phases, which confirms that most of the dominant user types have changed between the research phases.

\begin{table}[!ht]
\caption{Dominant user type}
\label{tab:hexad1}
\resizebox{0.5\textwidth}{!}{
\begin{tabular}{lll}
\hline
\textbf{User type}       & \textbf{1st}           & \textbf{2nd}         \\ \hline 

Philanthropist                                       & 24\%         & 23\%         \\ \hline
Achiever                                             & 12\%         & 16\%         \\ \hline
Free Spirit                                          & 5\%          & 8\%          \\ \hline
Socialiser                                           & 3\%          & 3\%          \\ \hline
Disruptor                                            & 3\%          & 3\%          \\ \hline
Player                                               & 1\%          & 4\%          \\ \hline
Achiever/Philanthropist                              & 20\%         & 8\%          \\ \hline
Achiever/Free Spirit/Philanthropist/Player/Socialiser & 4\%         & -            \\ \hline
Achiever/Free Spirit/Philanthropist                  & 4\%          & 4\%          \\ \hline
Free Spirit/Player                                   & 4\%          & 3\%          \\ \hline
Philanthropist/Player                                & 4\%          & 3\%          \\ \hline
Achiever/Player                                      & 3\%          & 1\%          \\ \hline
Achiever/Philanthropist/Player/Socialiser            & 3\%          & 3\%          \\ \hline
Achiever/Socialiser                                  & 3\%          & 3\%          \\ \hline
Achiever/Philanthropist/Socialiser                   & 1\%          & 1\%          \\ \hline
Philanthropist/Socialiser                            & 1\%          & 5\%          \\ \hline
Achiever/Free Spirit/Philanthropist/Player           & 1\%          & 3\%          \\ \hline
Achiever/Free Spirit                                 & 1\%          & 5\%          \\ \hline
Free Spirit/Philanthropist/Player/Socialiser         & 1\%          & -            \\ \hline
Achiever/Player/Socialiser                           & -            & 1\%          \\ \hline
Free Spirit/Philanthropist/Socialiser                & -            & 1\%          \\ \hline
Player/Socialiser                                    & -            & 1\%          \\ \hline

\multicolumn{3}{p{7cm}}{\textbf{Key}: 1st: First research phase; 2nd: Second research phase.} 
\\ \hline
\end{tabular}
}
\end{table}

Then, we calculated the average score for each user type in the research. Since each Hexad sub-scale is formed by four questions arranged in a 7-point Likert Scale, the maximum value a Hexad sub-scale can be is 28. Similar to other studies that accessed the user type through the Hexad scale \cite{tondello2016gamification,tondello2019empirical,altmeyer2020hexarcade}, the Philanthropists and Achievers presented the higher average score while the Disruptors presented the lower average score. After testing the normality of the data using the Shapiro-Wilk test, we measured the bivariate correlation coefficients using Kendall's $\tau$, since the user type scores were non-parametric. Considering the conversion table proposed by Gilpin \cite{gilpin1993table}, the Achievers' scores between the research phases presented a weak correlation while Socialisers', Free Spirits', Philanthropists', Disruptors', and Players' scores presented a moderate correlation. 

\autoref{tab:mean-scores} reports the average scores, the standard deviation, and the bivariate correlation coefficients (Kendall's $\tau$). These results indicate that, besides the differences in the dominant user types, the six Hexad sub-scales also presented differences in the average scores in both phases.

\begin{table}[!ht]
\caption{Mean scores, standard deviation, and bivariate correlation coefficients (Kendall's $\tau$)}
\label{tab:mean-scores}
\begin{tabular}{lccll}
\hline
\textbf{User Types} & \multicolumn{1}{l}{\textbf{Mean score}} & \multicolumn{1}{l}{\textbf{S.D.}} & \multicolumn{1}{c}{\textbf{$\tau$}} \\ \hline
Achiever1           & 24,69                                   & 4,03                              & \multirow{2}{*}{0,265**}       \\ \cline{1-3}
Achiever2           & 23,39                                   & 4,72                              &                                \\ \hline
Disruptor1          & 15,72                                   & 5,27                              & \multirow{2}{*}{0,434**}       \\ \cline{1-3}
Disruptor2          & 15,07                                   & 5,33                              &                                \\ \hline
Free Spirit1        & 23,42                                   & 3,98                              & \multirow{2}{*}{0,365**}       \\ \cline{1-3}
Free Spirit2        & 22,43                                   & 4,83                              &                                \\ \hline
Philanthropist1     & 24,92                                   & 4,07                              & \multirow{2}{*}{0,379**}       \\ \cline{1-3}
Philanthropist2     & 24,22                                   & 4,43                              &                                \\ \hline
Player1             & 21,18                                   & 4,99                              & \multirow{2}{*}{0,478**}       \\ \cline{1-3}
Player2             & 21,03                                   & 5,59                              &                                \\ \hline
Socialiser1         & 20,88                                   & 5,44                              & \multirow{2}{*}{0,358**}       \\ \cline{1-3}
Socialiser2         & 21,16                                   & 5,44                              &                                \\ \hline
\multicolumn{5}{p{6.5cm}}{\textbf{Key}: $\tau$: Kendall's tau; 1: results of the first research phase; 2: results of the second research phase; ** p<0.01; P: Philanthropist; A: Achiever; R: Player; F: Free Spirit; S: Socialiser; D: Disruptor.} 
\\ \hline

\end{tabular}
\end{table}

\subsection{Discussion}
\label{discussion}

This study examined if the Hexad's user types of 74 participants presented differences after six months. We analyzed the differences among the dominant user types and also the differences presented in the scores of the six Hexad sub-scales. Overall, our findings indicated that most of the participants presented changes in their dominant user types, and also the six Hexad sub-scales presented differences in the average scores after six months.

When we consider only the dominant user type (\textit{i.e.}, the strongest tendency), 76\% of the participants presented a change in the score. As showed in \autoref{tab:hexad1}, some dominant user types have presented a considerable change between the research phases (\textit{e.g.}, Achiever/Philanthropists were 20\% of the participants of the first research phase and decreased to 8\% in the second). Also is notable that in both phases, all the participants of some dominant user types have changed (\textit{e.g.}, Achiever/Player/Socialiser did not appear in the first research phase; Free Spirit/Philanthropist/Player/Socialiser appeared only in the first research phase). These results can be related to the results found by Busch \textit{et al}. \cite{busch2016player}, which indicated that, in the game context, some player types can be less stable over time. Our results indicate that \textbf{users present changes in the dominant user type (\textit{i.e.}, the strongest tendency) over time, and consequently, the dominant user types can not be considered stable.}

Our results also indicate that even when we consider only the dominant user type, users can present more than one of the basic Hexad user types (\textit{i.e.}, Achiever, Philanthropist, Socialiser, Free Spirit, Player, and Disruptor) as the dominant user type. As can be seen in \autoref{tab:hexad1}, in the first research phase there were more participants disposed in more than one user type (\textit{i.e.}, Achiever/Philanthropist = 20\%) than participants disposed in some of the basic Hexad user types (\textit{i.e.}, Achiever = 12\%, Free Spirit = 5\%, Socialiser = 3\%, Disruptor = 3\%, and Player = 1\%). As indicated by previous studies \cite{tondello2016gamification,tondello2019empirical}, some user types seem to be correlated, which might explain why some respondents presented more than one user type as the dominant user type. Since the user performance can be affected by the user type \cite{lopez2019effects}, we believe that, when considering only the dominant user type, \textbf{it is necessary to analyze if the user presents more than one user type as the dominant}. 

Considering the demographic information of the respondents that changed the dominant user type, women seem to be slightly more susceptible to change the dominant user type than men (see \autoref{results}). The study conducted by Tondello \textit{et al}. \cite{tondello2019empirical} showed that women scored higher in the intrinsically motivated user types (\textit{i.e.}, Achiever, Philanthropist, Socialiser, and Free Spirit), and since our results showed that the user types Achiever, Philanthropist, and Free Spirit presented the higher difference in the average scores of both research phases, we believe that \textbf{women might be more susceptible to change the user type than men because they are more motivated by the intrinsically motivated user types.} From the descriptive analysis, it was not possible to identify how age can influence the user types' change, since the changes varied in the age groups measured in this study (see \autoref{results}). Despite that, \textbf{our results demonstrate that people can change their user types in different life stages}, showing that to personalize gamification it is important to consider other characteristics besides the user type and age group of the users.

Even though the results about the differences presented between the user types in both research phases (see \autoref{tab:mean-scores}) can be considered small (all the user types presented less than 1.3 points of difference in the average scores, from 28 available), we understand that, when we consider all the average scores of the user types, the participants' changes might have produced an offsetting change. Thus, some participants might have increased a particular item while some decreased in the same item, therefore, producing offsetting changes. 

As outlined, the user types can not be considered stable after six months. Most of the participants have presented different dominant user types and also the results showed differences in the average scores of the user types in both phases. Our results demonstrate that when designing gamified environments based on user types it is necessary to measure the users' scores after a certain period. Personalization of gamified environments also needs to follow the user changes, guaranteeing that the personalization supports the user constantly.

\subsection{Limitations}
\label{limitations}

Our study has some limitations inherent to the type of study, which we seek to mitigate. Initially, the limited number of participants, as well as the fact that the participants are from the same country (\textit{i.e.}, Brazil) can prevent the generalization of the results. To mitigate this limitation we used consolidated static methods for validating responses. To access all the necessary information about the respondents we used a questionnaire that might be considered long (32 questions) for the respondents, thus, to mitigate this limitation, we conducted a pilot study asking the participants if they considered the size of the questionnaire adequate for the research. It is possible that some participants might have not paid attention when answering the survey. To mitigate this limitation, all the respondents were volunteers and we inserted an ``attention check question'', eliminating responses from participants who missed this question. Finally, since we only used descriptive statistics, a further statistical investigation is necessary to improve our results, especially the results based on the demographic information.  

\subsection{Research agenda}

Considering the obtained results and the limitations of this study, some possible studies can be conducted in the future. First, since this is an exploratory study and our results showed that the Hexad sub-scale for the Disruptor and Free Spirit presented reliability results slightly below the acceptable, and other studies have found similar results \cite{tondello2019empirical,ooge2020tailoring}, we believe this result might highlight \textbf{the necessity of the improvement of Disruptor and Free Spirit sub-scale or further analysis of these user types}. Also, the results highlight the \textbf{necessity of further investigation in each measurement Hexad item, psychometric modeling for the sets, and analysis of how the stability of personality traits can affect the changes of the user types.} 

This study was conducted without considering a specific domain, and the gamification effects can vary according to the context \cite{hallifax2019factors}. Literature reviews \cite{koivisto2019rise, klock2020tailored} have indicated before the necessity of studies in the gamification field that improve the understanding of the impact that the context presents in gamified settings. Thus, \textbf{we recommend that future studies replicate this study in specific contexts (\textit{e.g.}, health, education, business) to analyze how the context affects the user type changes, furthering our results.} 

Similar to other studies about player/user types \cite{busch2016player, tondello2019empirical}, we were able to access the user types of the respondents in different moments, and it was possible to identify that the number of participants tends to decrease when the research is conducted in more than one phase. One possible reason for the reduction of participants in our research is there was not any kind of intervention with the respondents between the study phases. Considering that the number of participants can reduce the exploratory power of the results \cite{koivisto2019rise}, and prevent the generalization of the results, we recommend that \textbf{future studies should be conducted with a larger sample and with interventions between the phases}.

Similar to Busch \textit{et al}. \cite{busch2016player}, we waited six months to analyze if the respondents have presented any difference in their user types. It is important to analyze if the participants can present changes in their user type earlier (\textit{e.g.}, after two or three months), as well as if the respondents can return to the first user type after a longer period (\textit{e.g.}, after a year). We suggest that \textbf{future studies should have more phases (\textit{e.g.}, two months, a year), considering the evaluation of the user types stability in a shorter and longer period}.

Our study identified there were differences between most of the respondents' user types after six months, confirming the hypothesis that the profile of people changes (after six months) and directly influencing the style of personalization that should be carried out (\textit{i.e.}, demonstrating the importance of dynamic personalization). Although recent studies \cite{altmeyer2019towards,altmeyer2020hexarcade} showed that the prediction of the Hexad user types might be a possibility, the user type is still mostly accessed through surveys and questionnaires \cite{klock2020tailored}. Considering that our results imply having to constantly analyze the users' type, making the process more costly by making users need to answer the questionnaire constantly, \textbf{future studies should be done to try to predict people's user type based on interaction data, machine learning, or based on the questionnaire response the first time (\textit{i.e.}, predict what the user will be after a certain time, based on questionnaire responses only once)}.

In \autoref{tab:research-agenda-summary} we summarize the research agenda.

\begin{table*}[ht]
\caption{Research agenda summary}
\label{tab:research-agenda-summary}
\begin{tabular}{p{5cm}p{5cm}p{3cm}}
\hline
\textbf{Recommendation}                                                        & \textbf{Motivation}                                                                            & \textbf{Type of studies}                      \\ \hline
Improvement of Disruptor and Free Spirit sub-scale       &Increase the reliability results      &Exploratory/surveys/factor analysis
 \\ \hline
Investigation of each measurement Hexad item & Analyze which Hexad item affects more the user changes & Exploratory/surveys     
 \\ \hline
Investigation about the stability of personality traits in the respondents &Analyze how the stability of personality traits affects the stability of the user type &Exploratory/surveys 
 \\ \hline 
Replicate the study in different areas                                        & Analyze whether contexts affect the stability of the user type                                                    & Exploratory/surveys                           \\ \hline
Replicate the study increasing the number of participants and interventions & Increase the reliability results and generalization power                                       & Exploratory/surveys                           \\ \hline
Predict how user types have changed over time                               & Avoid the need to repeatedly apply questionnaires and facilitate the personalization process & Exploratory/surveys/machine learning analysis \\ \hline
\end{tabular}
\end{table*}

\section{Concluding remarks}
\label{concluding-remarks}

In this study, divided into two different phases, we conducted a comparison of 74 people's user types, to analyze if they can be considered stable. We used the Gamification User Types Hexad to access the participants' user types (Achiever, Philanthropist, Socialiser, Free Spirit, Player, and Disruptor) and conducted the second research phase after six months of the first phase. Our main results showed that the dominant user type of most of the participants has changed over time. Also, when comparing the average scores of the participants' user types in both phases, the average scores were different, indicating that changes happened between the phases. These results demonstrate that when designing a gamified environment based on the user type, it is important that this design can support user type's changes since the user types can not be considered stable. As future studies, we aim to focus on measuring the user types changes after a different period (\textit{i.e.}, a year) and with people with a different demographic background (\textit{i.e.}, more than one country). Also, since this study was exploratory, we intend to do further statistic analysis (\textit{e.g.}, by using Structural Equation Modeling) about all the variables of each Hexad sub-scale and work in predict the user changes from the previously collected information.

\begin{acknowledgments}
  The authors would like to thank the grant provided by São Paulo Research Foundation (FAPESP), Projects: 2018/07688-1 and 2020/02801-4.
\end{acknowledgments}

\end{document}